\documentclass{article}
\usepackage{graphicx} 
\usepackage[a4paper,top=25mm,bottom=25mm,left=25mm,right=25mm,nohead,nofoot,includeheadfoot]{geometry}
\usepackage{indentfirst}
\usepackage{multicol}
\usepackage{caption}
\usepackage{hyperref}
\usepackage[labelfont=bf]{caption}
\usepackage{amsmath}
\usepackage{amssymb}
\usepackage{physics}
\usepackage{float}
\usepackage{lipsum}
\usepackage[backend=biber, style=phys, sortcites=true, biblabel=brackets]{biblatex}
\begin{document}
\twocolumn[
\begin{center}
{\LARGE \bfseries
Sparsity-Driven Entanglement Detection in High-Dimensional Quantum States \par}
\vspace{1em}
{\large\textbf{
Stav Lotan\textsuperscript{1,2,3}, Hugo Defienne\textsuperscript{4}, Ronen Talmon\textsuperscript{1}, and Guy Bartal\textsuperscript{1,2,3,*} }\par}
\vspace{1em}

{\footnotesize
\textsuperscript{1}Andrew and Erna Viterbi Department of Electrical \& Computer Engineering, Technion - Israel Institute of Technology, Haifa 32000, Israel \\
\textsuperscript{2}Russell Berrie Nanotechnology Institute, Technion - Israel Institute of Technology, Haifa 32000, Israel \\
\textsuperscript{3}Helen Diller Quantum Center, Technion - Israel Institute of Technology, Haifa 32000, Israel \\
\textsuperscript{4}Sorbonne Universit\'e, CNRS, Institut des NanoSciences de Paris (INSP), F-75005 Paris, France \\
\textsuperscript{*}Corresponding author
}
\end{center}

\begin{center}
\begin{minipage}{0.9\linewidth}
\setlength{\parindent}{2em}
\begin{abstract}
The characterization of high-dimensional quantum entanglement is crucial for advanced quantum computing and quantum information algorithms. Traditional methods require extensive data acquisition and suffer from limited visibility due to experimental noise. Here, we introduce a sparsity-driven framework to enhance the detection and certification of high-dimensional entanglement in spatially entangled photon pairs. By applying $\ell_1$-regularized reconstruction to sample covariance matrices obtained from measurements on photons produced via spontaneous parametric down-conversion (SPDC) measurements, we enhance the visibility of the correlation signal while suppressing noise. We demonstrate, using a position-momentum Einstein-Podolsky-Rosen (EPR) entanglement criterion, that this approach enables certification of an entanglement dimensionality that cannot be achieved without regularization. Our method is scalable, simple to use and compatible with existing quantum-optics platforms, thus paves the way for efficient, real-time analysis of high-dimensional quantum states.
\end{abstract}
\end{minipage}
\end{center}

\vspace{2em}
]

\section*{Introduction}
Quantum entanglement lies at the heart of many quantum technologies, underpinning advances in secure communication \cite{paraiso_photonic_2021}, quantum computing \cite{obrien_optical_2007}, and enhanced sensing \cite{defienne_advances_2024}. Traditionally, entanglement has been studied and exploited in low-dimensional systems, often involving two-level qubits, where both theory and experimental techniques are mature. However, bipartite entanglement in low dimensions suffers from fundamental limitations, including low information capacity, vulnerability to noise \cite{cozzolino_high-dimensional_2019}, and restricted scalability for real-world quantum protocols\cite{brock_quantum_2025}.

High-dimensional entanglement, where entanglement is distributed across larger Hilbert spaces such as spatial \cite{ndagano_imaging_2020}, temporal \cite{cheng_high-dimensional_2023} or orbital angular momentum modes \cite{fickler_interface_2014}, provides a promising path to overcome these limitations ~\cite{erhard_advances_2020, armstrong_programmable_2012}. It allows more information to be encoded per photon pair \cite{ding_high-dimensional_2017}, increases noise resilience \cite{islam_provably_2017}, and enhances quantum computing schemes \cite{kues_-chip_2017}. These properties make high-dimensional entangled states especially attractive for tasks like high-rate quantum key distribution \cite{cozzolino_high-dimensional_2019}, quantum imaging \cite{defienne_quantum_2019}, and error correction \cite{brock_quantum_2025}.

However, characterizing and certifying high dimensional entanglement remains a challenge. The large number of modes involved leads to an exponential growth in the number of measurement data needed, and the signal associated with each entangled mode is typically weak. Moreover, practical single-photon sensitive sensors, such as Electron multiplying Charge Couple Device (EMCCD) camera or single-photon avalanche diode (SPAD) cameras, suffer from dark counts, low fill factors, and limited temporal resolution, all of which introduce noise that obscures the underlying quantum correlations~\cite{madonini_single_2021,lubin_photon_2022}. In addition, noise arises from electronic readout, stray light, and especially accidental coincidence detections of multi-photon pair events within the same frame, which need to be subtracted. As a result, extracting a clear correlation map or certifying high-dimension entanglement often requires over tens of millions of frames, which can take hours or days of acquisition time~\cite{ndagano_imaging_2020, defienne_general_2018, courme_quantifying_2023,moreau_realization_2012,edgar_imaging_2012}.

Previous research has made significant strides in addressing these challenges. Efforts have included optimizing physical imaging systems~\cite{rambach_robust_2021}, characterizing sensor noise to extract correlations~\cite{defienne_general_2018, reichert_optimizing_2018}, and implementing advanced sensors such as SPAD arrays and recent time-stamping cameras to increase detection rates~\cite{ndagano_imaging_2020, defienne_full-field_2021,courme_quantifying_2023}. 
Those approaches focus on acquisition hardware or raw sensitivity. In contrast, the algorithms used to process and interpret the raw data, such as sample covariance calculations or $g^{(2)}$ correlation, have remained largely unchanged and often lead to noisy outputs that obscure the signal of interest.

The structure of high-dimensional entangled states offers an opportunity: these states are often sparse in the measurement basis [see supplementary A]. This natural sparsity suggests that tools from machine learning and statistical signal processing, particularly matrix recovery techniques with $\ell_1$ regularization \cite{candes_enhancing_2008}. This regularization promotes sparsity by penalizing the absolute sum of matrix elements, effectively shrinking noisy values toward zero while preserving strong and meaningful correlations. Similar tools have proven effective in compressive sensing and structured signal for quantum imaging and quantum state tomography for low-rank density matrix \cite{oren_sparsity-based_2016, katz_compressive_2009, liu_experimental_2012, jiying_high-quality_2010, gross_quantum_2010, flammia_quantum_2012,howland_efficient_2013,schneeloch_quantifying_2019}.

Here, we present a robust method to characterize high-dimensional quantum states under noisy conditions by employing statistical signal processing tools. We extract the correlation matrix and the entanglement dimension witness from noisy high-dimensional quantum data by utilizing the sparse nature of the quantum state density matrix, and show that it suppresses noise and recovers structured features that would otherwise be obscured. By applying $\ell_1$-regularized reconstruction to sample covariance matrices derived from Spontaneous Parametric Down-Conversion (SPDC) experiments, we further show that this enhanced visibility directly translates into improved entanglement certification, increasing the lower bound of the entanglement dimension certification by nearly 150\% when using the SPAD camera. Furthermore, when using more noisy sensors such as the EMCCD camera, it can be the difference between violating the EPR or not using the same dataset, and without additional measurements.

\section*{Results}

We evaluate our sparsity-based reconstruction method using two types of sensors: an Andor iXon Ultra 888 EMCCD camera, which has high resolution but has limited frame rate and therefore is more vulnerable to accidental coincidence detections of multi-photon pair events within the same frame, and a Micro Photon Devices SPC3 SPAD camera (data extracted from \cite{ndagano_imaging_2020}) which has higher frame rate but suffers from low detection rate and low resolution \cite{lotan_leveraging_2025}.
In both setups, spatially entangled photon pairs were generated via SPDC using a type-I Beta-Barium Borate (BBO) crystal pumped by a $404nm$ continuous-wave laser.

\begin{figure}[ht]
\centering
\includegraphics[width=0.95\linewidth]{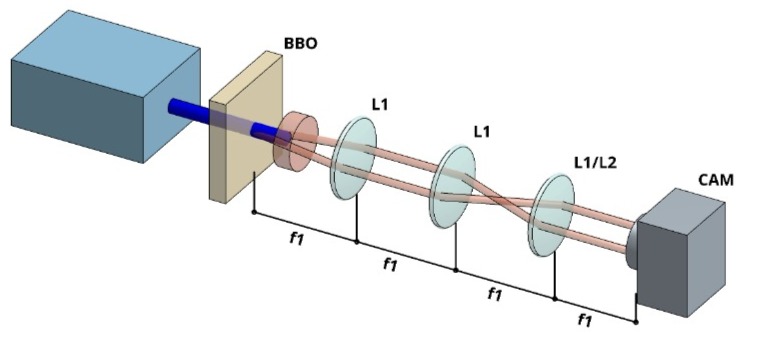}
\caption{Schematic of the experimental setup. Spatially entangled photon pairs are generated via SPDC and imaged in both imaging, \textbf{$L_2$}, and Fourier, \textbf{$L_1$}, configurations using an EMCCD camera. Here, $f_1 = 100[mm]$ and $f_2 = 50[mm]$ are the focal length of $L_1$ and $L_2$ respectively.}
\label{fig:setup}
\end{figure}

For each camera, two optical configurations were implemented to record photon correlations: an imaging configuration to access position correlations, and a Fourier configuration to access momentum correlations. The acquired intensity images were used to compute the sample covariance matrices \cite{defienne_general_2018}:

\begin{equation}
\mathrm{\Sigma_0}(i, j)
= \langle I_i I_j \rangle \;-\; \langle I_i \rangle \, \langle I_j \rangle,\label{eq:covariance}
\end{equation}
{\setlength{\parindent}{0cm}
where $I_i$ and $I_j$ are the intensity measurements at pixels $i$ and $j$, and $\langle \cdot \rangle$ denotes an average.
}
To perform the sparse reconstruction, we define an optimization problem that combines fidelity to the measured sample covariance matrix $\Sigma_0$ with a sparsity-promoting penalty \cite{candes_enhancing_2008}. Specifically, we solve the following objective function for EMCCD:

\begin{equation}
  \Sigma^{*} = \arg\min_{\Sigma} \left\{\lVert \Sigma - \Sigma_{0} \rVert
      + \lambda\,\lVert \Sigma + |\Sigma_{min}| \rVert_{1}\right\}
  \label{eq:objective}
\end{equation}
{\setlength{\parindent}{0cm}
where $\Sigma_0$ is the measured sample covariance matrix, $||\Sigma - \Sigma_0||$ is the Frobenius norm (i.e., mean squared error) between the reconstruction and the raw data, $\rVert \Sigma + |\Sigma_{min}| \rVert_1$ is the sum of the absolute values of the matrix elements (the $\ell_1$ norm) shifted by the minimum value of $\Sigma$, $|\Sigma_{min}|$, and $\lambda$ is a regularization hyperparameter that controls the trade-off between data fidelity and sparsity. The SPAD camera measurements did not require such a shift, as the statistical noise after the subtraction of the accidental coincidences is close to zero.
}
This formulation suppresses noise and small-valued artifacts in $\Sigma_0$, while retaining the most significant information, the correlated components of the matrix. We note that the objective function does not use any prior knowledge or assumptions about the matrix rank or the quantum state, except for sparsity.

We first analyze the signal-to-noise ratio (SNR) \cite{reichert_optimizing_2018} on the EMCCD camera data as a function of the number of images used in the reconstruction of the sample covariance. The SNR is defined as:

\begin{center}
\begin{equation}
SNR = \frac{\mu_{\text{peak}}}{\sigma_{\text{background}}}
\end{equation}
\end{center}
{\setlength{\parindent}{0cm}
where $\mu_{\text{peak}}$ is the maximum value of the peak area in the sum/minus coordinate projections of the covariance matrix, and $\sigma_{\text{background}}$ is the standard deviation of the background in the sum/minus coordinate projections of the covariance matrix, see supplementary B.}

\begin{figure}[ht]
\centering
\includegraphics[width=\linewidth]{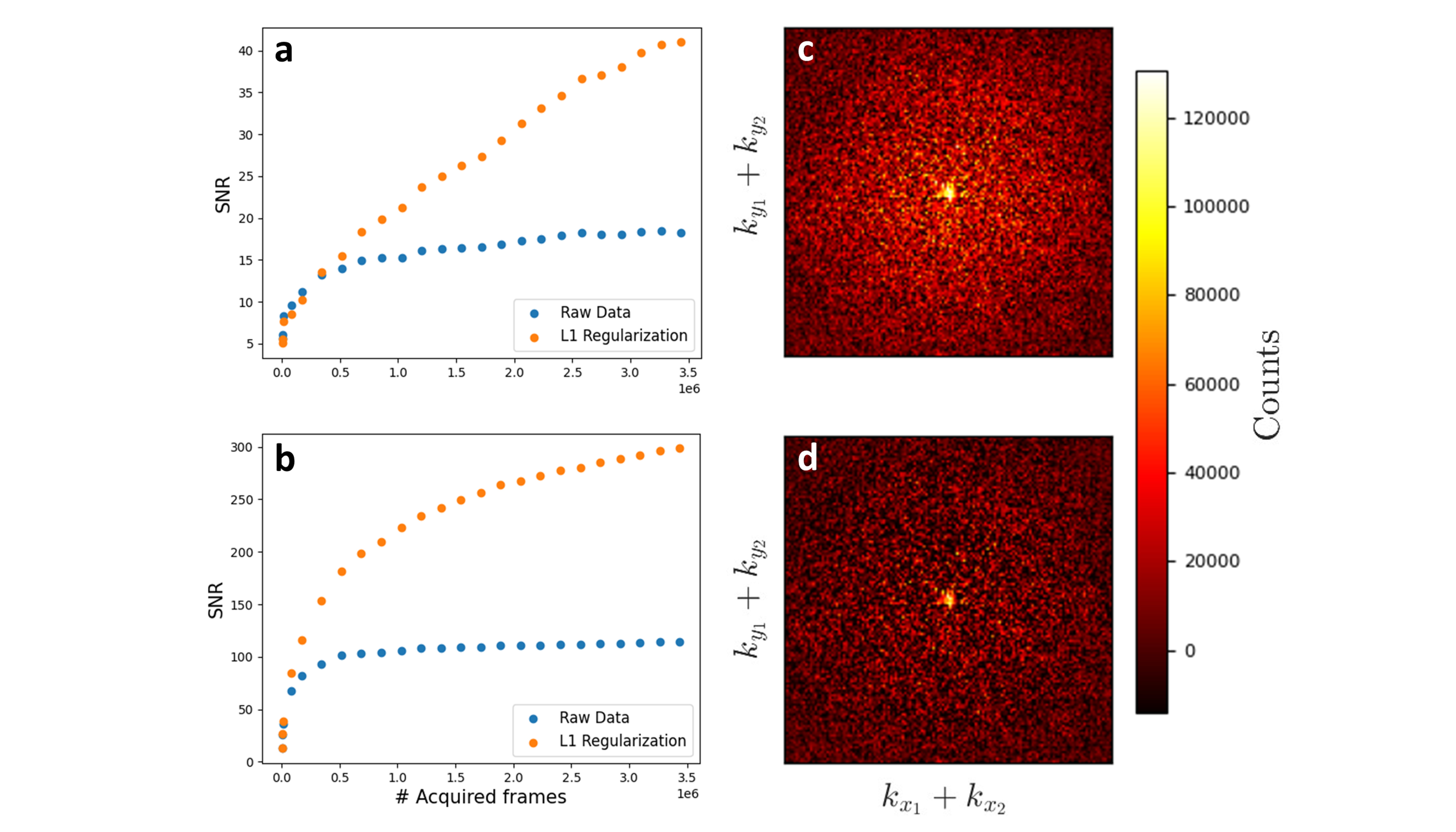}
\caption{EMCCD SNR as a function of the number of frames acquired for unoptimized and optimized sample covariance matrices, \textbf{(a)} Sum-coordinate projection of the covariance matrix \textbf{(b)} Minus-coordinate projection of the covariance matrix. \textbf{(c-d)} are the Sum-coordinate projection of the covariance matrix using $\sim10^5$ frames: \textbf{c} unoptimize data and \textbf{d} optimize data.}
\label{fig:SNR_EMCCD}
\end{figure}

We compare the SNR obtained from the EMCCD raw data to the data obtained after applying our $\ell_1$-based sparse reconstruction algorithm, as shown in Figure~\ref{fig:SNR_EMCCD}. Evidently, the optimized matrices show significantly higher SNR despite the relatively small number of frames ($\sim10^5$), where the sparse reconstruction reveals structured correlation patterns that are otherwise obscured in the raw data. 

We repeat the same analysis for the SPAD camera, evaluating the SNR of the sample covariance computed from $10^7$ frames. To highlight the emergence of the quantum structure after optimization, we extract the conditional image, defined as the correlation between a single reference pixel and the entire frame, and display it on a logarithmic scale. The optimization strongly suppresses background noise and enhances the visibility of the quantum correlation.

\begin{figure}[ht]
\centering
\includegraphics[width=1\linewidth]{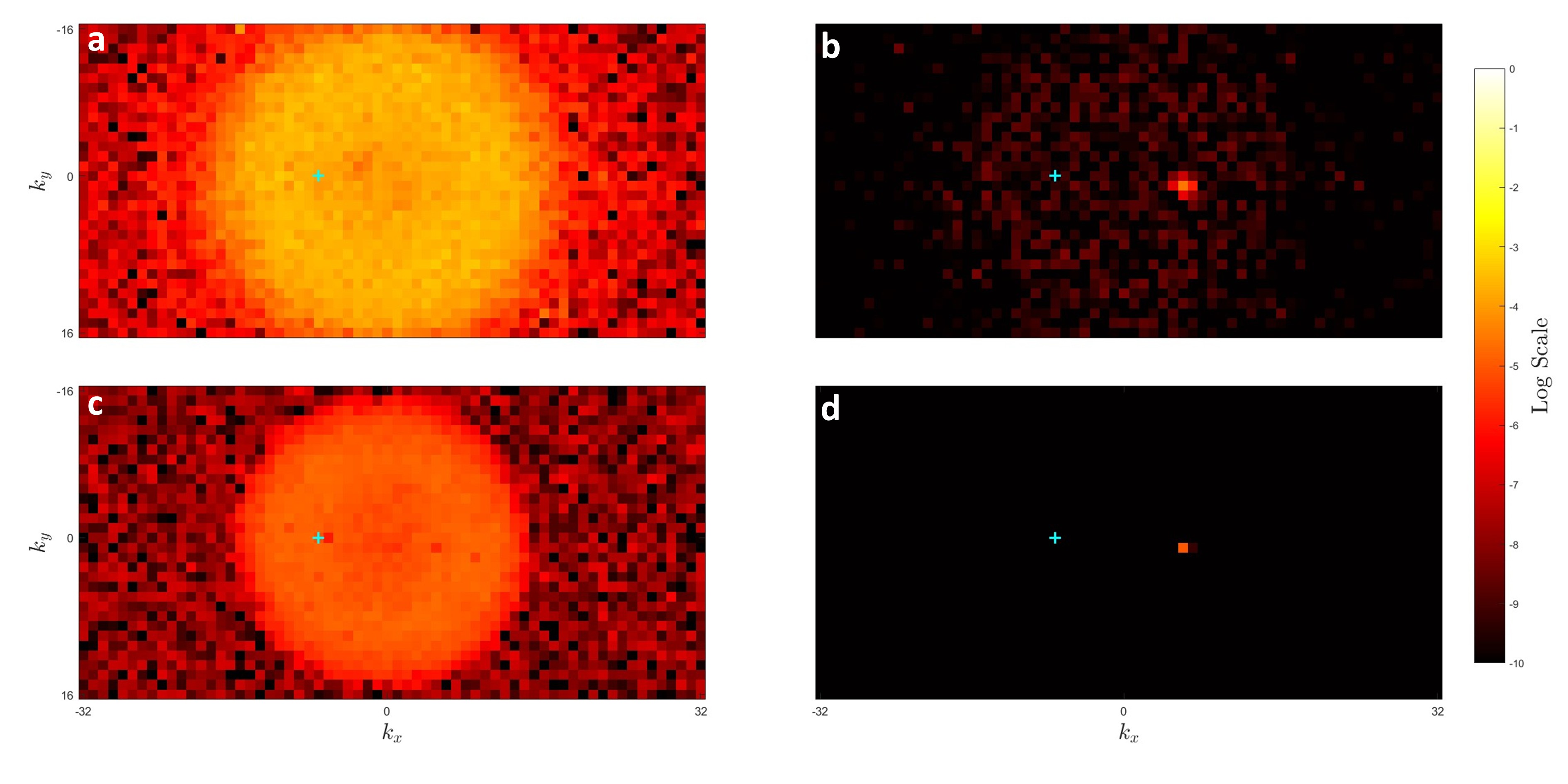}
\caption{SPAD camera results in Fourier basis after $10^7$ frames in a logarithmic scale. \textbf{(a,c)} Unoptimized full covariance matrix and conditional image (correlation of a single exemplary pixel to the whole frame), both reshaped into an image. \textbf{(b,d)} Same as \textbf{(a,c)} but optimized using Eq.~\ref{eq:objective}.}
\label{fig:SNR_SPAD}
\end{figure}

\begin{figure}[ht]
\centering
\includegraphics[width=\linewidth]{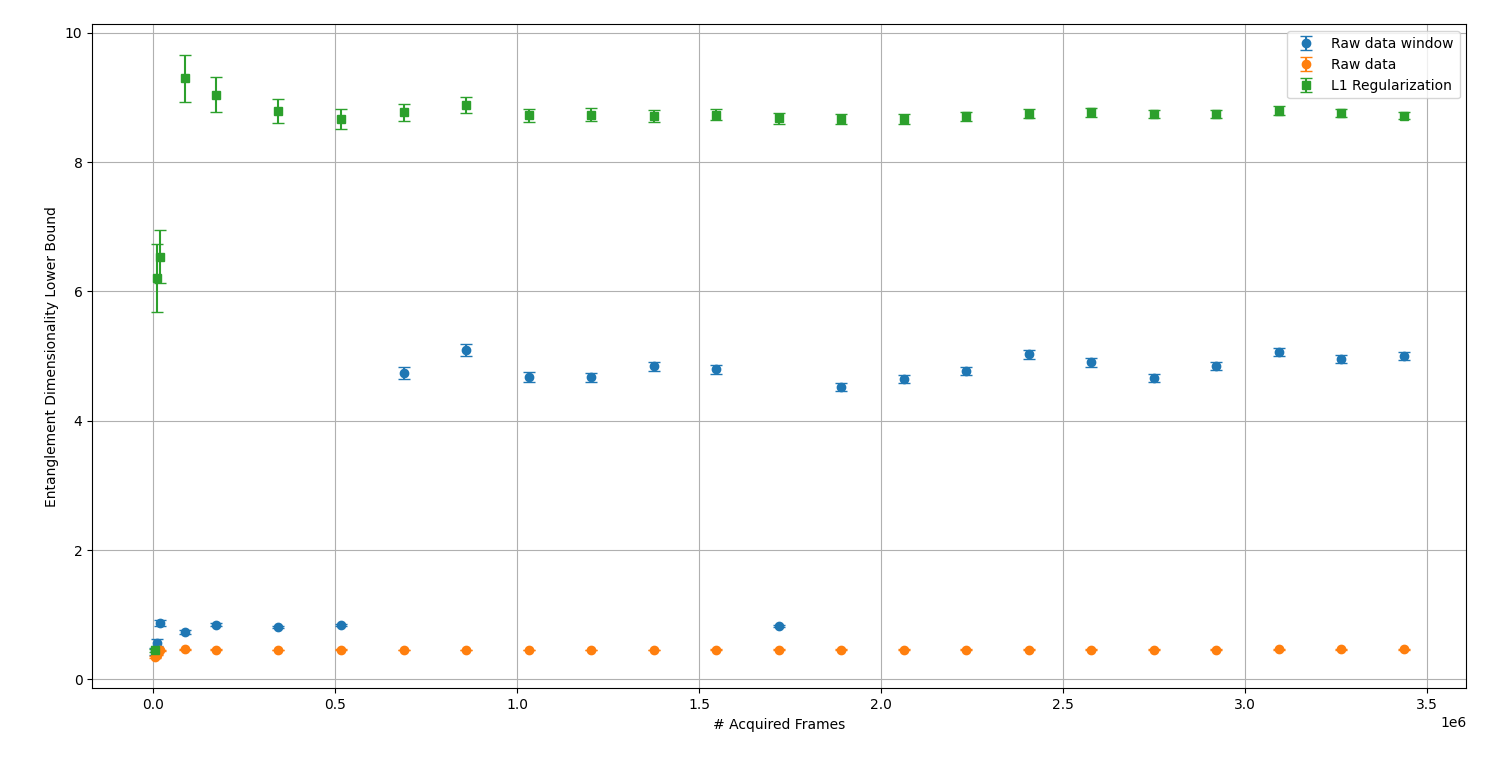}
\caption{Entanglement dimension witness calculation as a function of the number of images.}
\label{fig:dimension}
\end{figure}

To quantify the impact of our method on entanglement detection, we apply an entanglement dimensionality witness based on the Einstein–Podolsky –Rosen (EPR) violation criterion and the assumption of a Double-Gaussian model~\cite{schneeloch_quantifying_2018,schneeloch_introduction_2016,reid_demonstration_1989}. The witness is calculated using the following equation:

\begin{equation}
-\log_{2}\left(e\,\Delta x_{minus}\,\Delta k_{sum}\right) = d
  \label{eq:ED}
\end{equation}
{\setlength{\parindent}{0cm}
where $\Delta x_{minus}$ and $\Delta k_{sum}$ denote the uncertainties extracted from the minus and sum coordinate projections of the covariance matrix, and $d$ value use to produce the entanglement dimension lower bound of $2^{d}$.} Figure \ref{fig:dimension} depicts the Entanglement dimension witness as a function of the number of acquired frames for three EMCCD cases: raw data, reduced Hilbert space (windowed) raw data, and optimized data obtained using Eq.~\ref{eq:ED}.

The raw data do not violate the EPR criterion, this happens due to gain and pump power fluctuations which causes the background of the sum/minus coordinate projection to have a non-flat shape \cite{defienne_general_2018}, that leads to a poor Gaussian fit, and saturate at a value of 0.8, which is equivalent to an entanglement dimension lower bound of 1, (equation \ref{eq:ED}). By reducing the Hilbert space by almost 4 times, using a window of $50\times50$ pixels instead of the full image $90\times90$ pixels, we obtain an EPR violation and an entanglement dimension lower bound of 5. In contrast, applying our sparse reconstruction method achieves a violation after only $\sim 2000$ frames and certifies a lower bound to the entanglement dimensionality of 9 using the same dataset. Moreover, the optimized data exhibit much faster convergence, background noise is strongly suppressed and nearly vanishes after $10^5$ frames, while the correlation peak remains unchanged (see Supplementary B).
In the SPAD data, when comparing the optimized and raw sample covariances that were reconstructed from $10^7$ frames, we observe an improvement of almost 150\%, from 15 to 21, in the entanglement dimension witness. 

\begin{figure}[ht]
\centering
\includegraphics[width=\linewidth]{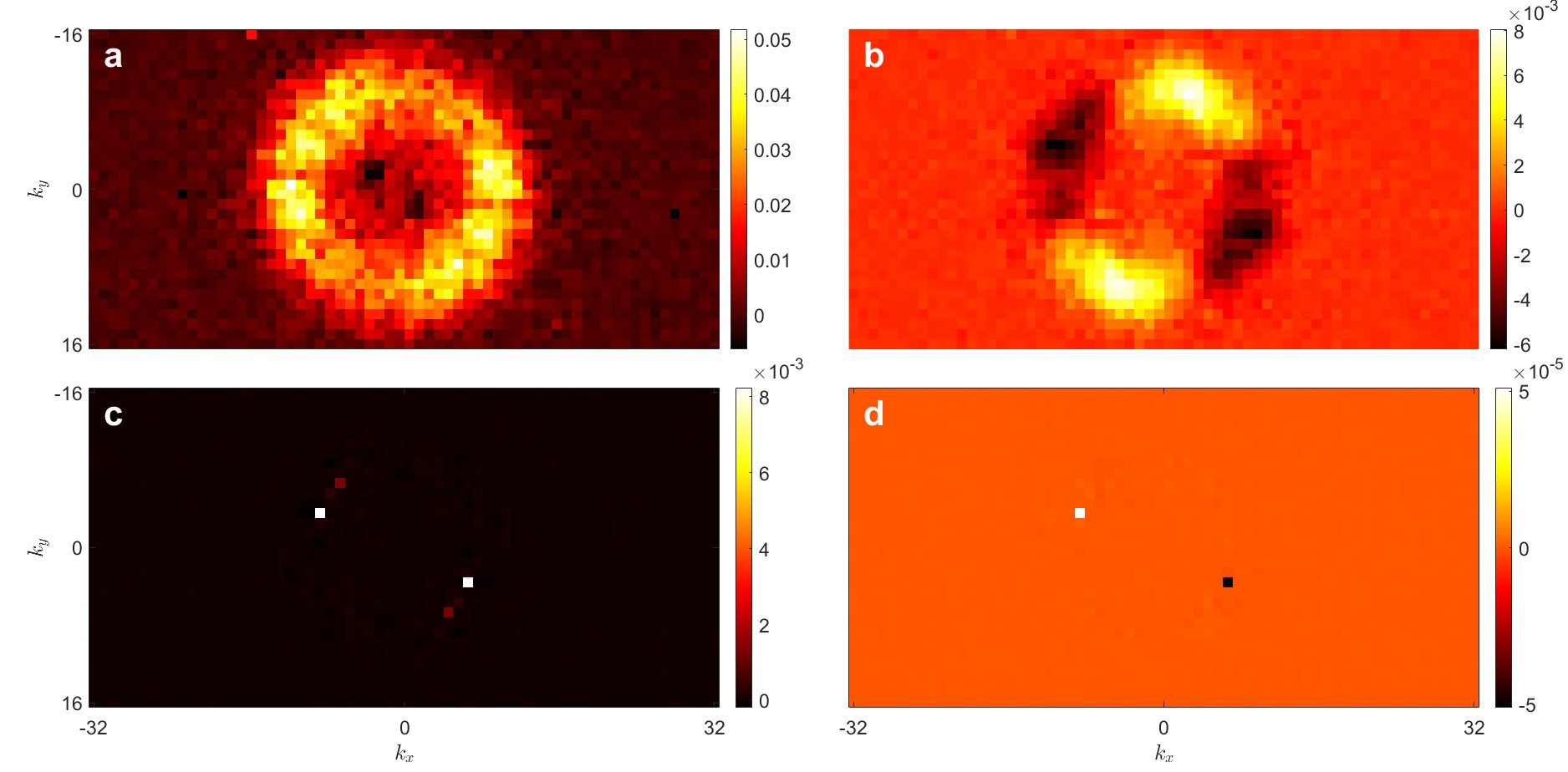}
\caption{The dominant eigenvectors from spectral analysis of the sample covariance matrices reshaped into image form. \textbf{(a)} \& \textbf{(b)} first and second eigenvectors of the raw data multiply by their compatible eigenvalues in absolute value and reshaped into image form. \textbf{(c)} \& \textbf{(d)} are the same for the optimized matrix.}
\label{fig:spectrum}
\end{figure}

Finally, we analyze the modal content of the data by performing a spectral decomposition on the covariance matrices reconstructed from the SPAD camera measurements as shown in equation \ref{eq:spectral}.

\begin{equation}
\Sigma_0 = \sum_{i=1}^{N} \lambda_i \, \mathbf{v}_i \mathbf{v}_i^{\mathsf{T}},
\label{eq:spectral}
\end{equation}
{\setlength{\parindent}{0cm}
where $\lambda_i$ and $\mathbf{v}_i$ are the eigenvalues and eigenvectors, respectively, and $\Sigma_0$ is the sample covariance matrix.} For an ideal bi-photon state, the full density matrix is expected to be diagonal, yielding a one-hot representation for the eigenvectors, where each eigenvector $\ket{v_i} = (0_1, \dots, 1_i, \dots, 0_n)$ corresponds to correlation between exactly one pair of pixels. Under the assumption that only the main diagonal of the density matrix has nonzero elements, the spectral decompositions of the covariance and density matrices can be comparable when the phase is assumed equal for all eigenvalues. We identify the two dominant eigenvalues and their associated eigenvectors for the unprocessed and optimized data, and present the reshaped eigenvectors as images in Figure~\ref{fig:spectrum}. In the raw data, the eigenvectors appear delocalize, where any single pixel appears to correlate broadly across the image. In contrast, the eigenvectors of the optimized matrix are spatially localized and exhibit a clear pairwise structure, consistent with the expected sparsity of the underlying entangled state. This decomposition not only provides insight into the effective dimensionality of the state but also reinforces the suitability of using sparsity-based techniques to extract meaningful structure from experimental data.

\section*{Conclusion}
Our results demonstrate that leveraging sparsity in high-dimensional quantum systems provides a powerful route to enhancing entanglement detection and certification. By applying $\ell_1$-regularized reconstruction to experimentally measured covariance matrices, we are able to significantly increase visibility, reveal meaningful structure with fewer measurements, and certify a higher entanglement dimensionality.


Importantly, this method operates entirely at the data-processing level, requiring no modifications to the experimental setup or acquisition protocol. It is compatible with both SPAD and EMCCD cameras, which are highly used in quantum state characterization and can be compatible with every other sensor. This generality indicates that sparsity-based techniques could be broadly applied to other forms of entanglement, including time-bin, polarization, or hybrid encoding schemes.

The enhanced visibility achieved with fewer frames also opens the door to faster or real-time entanglement certification. This is particularly relevant for practical quantum communication and imaging applications, where long integration times limit scalability. 

Furthermore, the improved spectral localization observed in the eigenmode decomposition provides strong evidence that the reconstructed covariance matrices retain physically meaningful structure. In particular, the emergence of localized or one-hot-like eigenmodes aligns with theoretical expectations for ideal bi-photon high-dimensional entangled states. This spectral behavior serves as a valuable consistency check, confirming that the reconstructed state is not only mathematically sparse but also physically plausible.

Overall, our findings highlight that incorporating prior knowledge of the quantum state structure, in this case sparsity, into the quantum data analysis can significantly improve performance, enabling more efficient and robust characterization of high-dimensional quantum systems.

\section*{Methods}
\subsection*{SPDC Sparsity}


Sparsity refers to the property of a signal or matrix in which only a small subset of elements carries significant information, while the rest are zero or negligibly small. In the context of high-dimensional quantum optics experiments, such as SPDC, this is achieved due to phase-matching conditions, which cause the bi-photon state to have strong correlations in both the momentum and position bases. 
In the momentum basis, the conservation of transverse momentum imposes that $k_1 + k_2 \approx 0$, while in the position basis, the photons tend to be emitted in the same spatial region due to the shared origin of the nonlinear interaction, resulting in $x_1 \approx x_2$. Therefore, the two-photon wavefunction is narrowly localized along the anti-diagonal in the momentum space and along the diagonal in the position space. As a result, the ideal covariance matrices derived from joint detections in either basis are expected to be sparse, with non-negligible values concentrated along these narrow correlation axes. 

For example, when calculating the sparsity of the SPAD correlation matrix in the position basis, we expect the following ratio:

\begin{align}    
    S &= 100 \cdot \left(1 - \frac{1}{\text{dim}_x \cdot \text{dim}_y} \right) =\\
    &= 100 \cdot \left(1 - \frac{1}{64 \cdot 32} \right) \approx 99.95\%
    \label{eq:sparse_calc}
\end{align}

Here, $S$ is the percentage of sparsity in the covariance matrix, $\text{dim}_x$ and $\text{dim}_y$ are the dimensions of the detection region in camera pixels. This result shows that only a fraction of the matrix values represent the quantum data, which we expect to see in SPDC process. 

Our experimental results validate this sparsity. As shown in Figures~\ref{fig:EMCCD_K} and~\ref{fig:EMCCD_P}, the raw covariance matrices exhibit narrow correlation structures embedded in a background of noise. In both bases, the true signal occupies a small fraction of the total pixel-pixel space, confirming the underlying sparsity of SPDC correlations in our detection basis.

\subsection*{Dimension witness extraction}
\subsubsection*{Sum/Minus coordinate projections of the covariance matrix}
Given detection coordinates $(x_1, x_2)$ in position basis or $(k_1, k_2)$ in momentum basis, we define the sum and minus coordinate projections of the covariance matrix as:

\begin{equation}
k_{\text{sum}} = {k_1 + k_2}, \quad x_{\text{minus}} = {x_1 - x_2}
\label{eq:sum_difference}
\end{equation}

\begin{figure}[H]
\centering
\includegraphics[width=\linewidth]{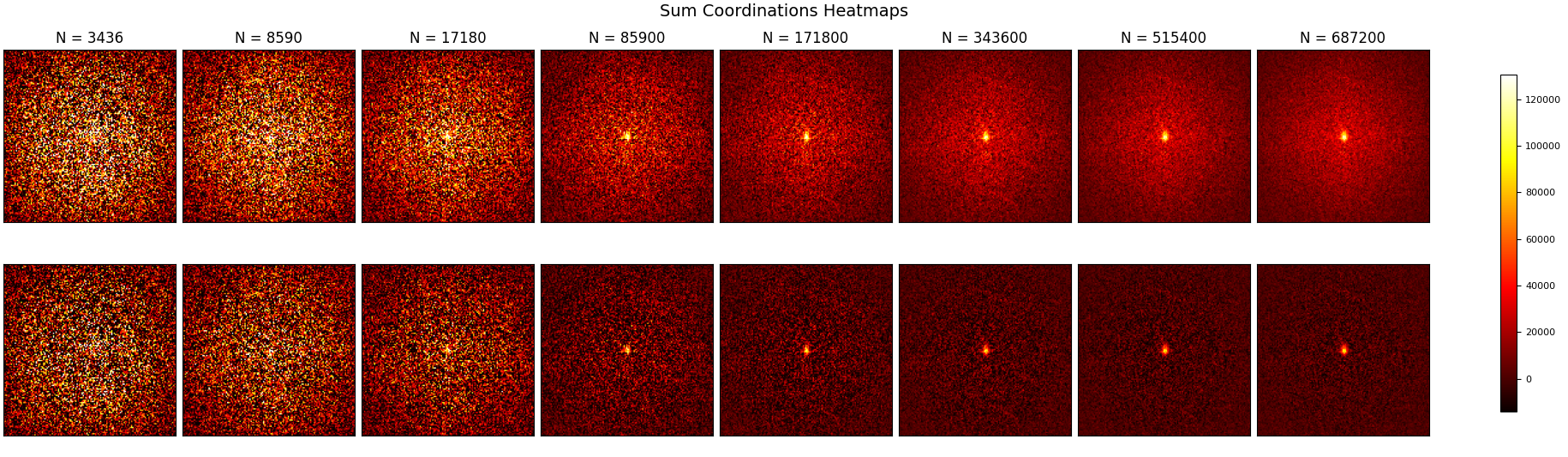}
\caption{Sum-coordinate projection of the covariance matrix plots for varying numbers of acquired frames. Top: raw data; bottom: $l_1$-regularized optimized data.}
\label{fig:EMCCD_K}
\end{figure}

\begin{figure}[H]
\centering
\includegraphics[width=\linewidth]{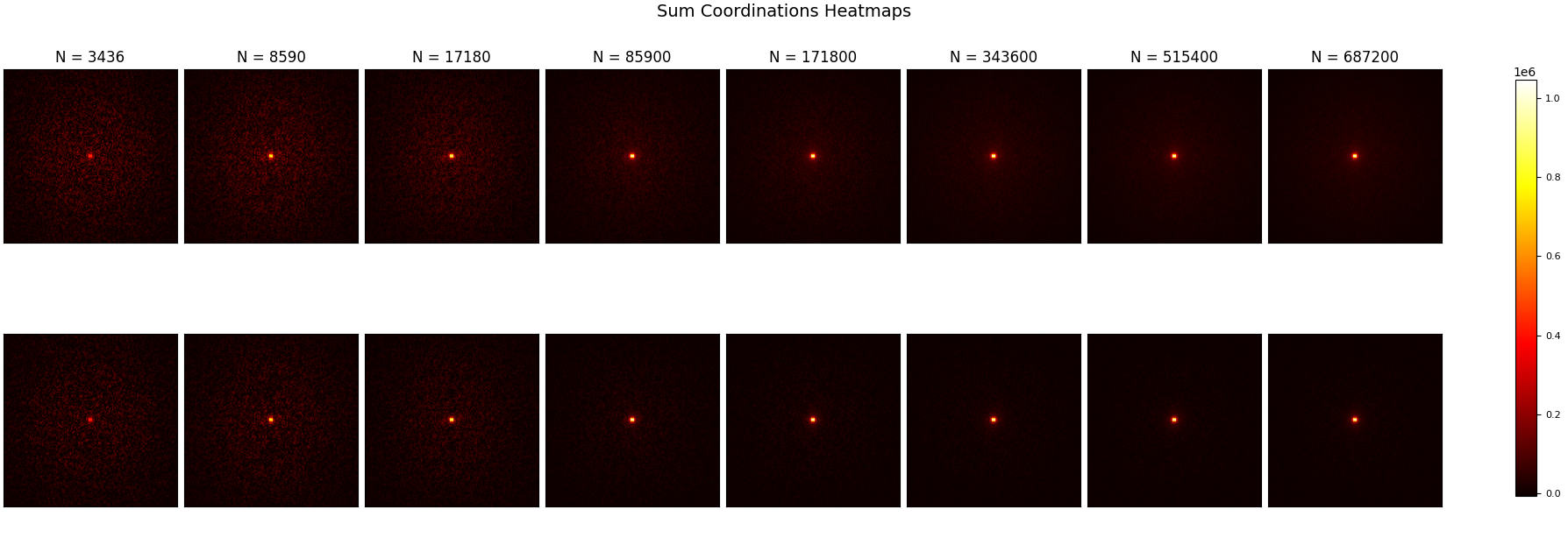}
\caption{Minus-coordinate projection of the covariance matrix plots for varying numbers of acquired frames. Top: raw data; bottom: $l_1$-regularized optimized data.}
\label{fig:EMCCD_P}
\end{figure}

We present the results of applying sum and minus coordinate projections of the covariance matrix in Figures \ref{fig:EMCCD_K} and \ref{fig:EMCCD_P}. The $l_1$-regularized optimization effectively suppresses spurious signals originating from multiple photon-pair events in the same EMCCD frame. In addition, the signal peak becomes distinguishable after approximately $10^5$ frames.

\subsubsection*{Correlation Width Extraction}

\begin{figure}[H]
\centering
\includegraphics[width=\linewidth]{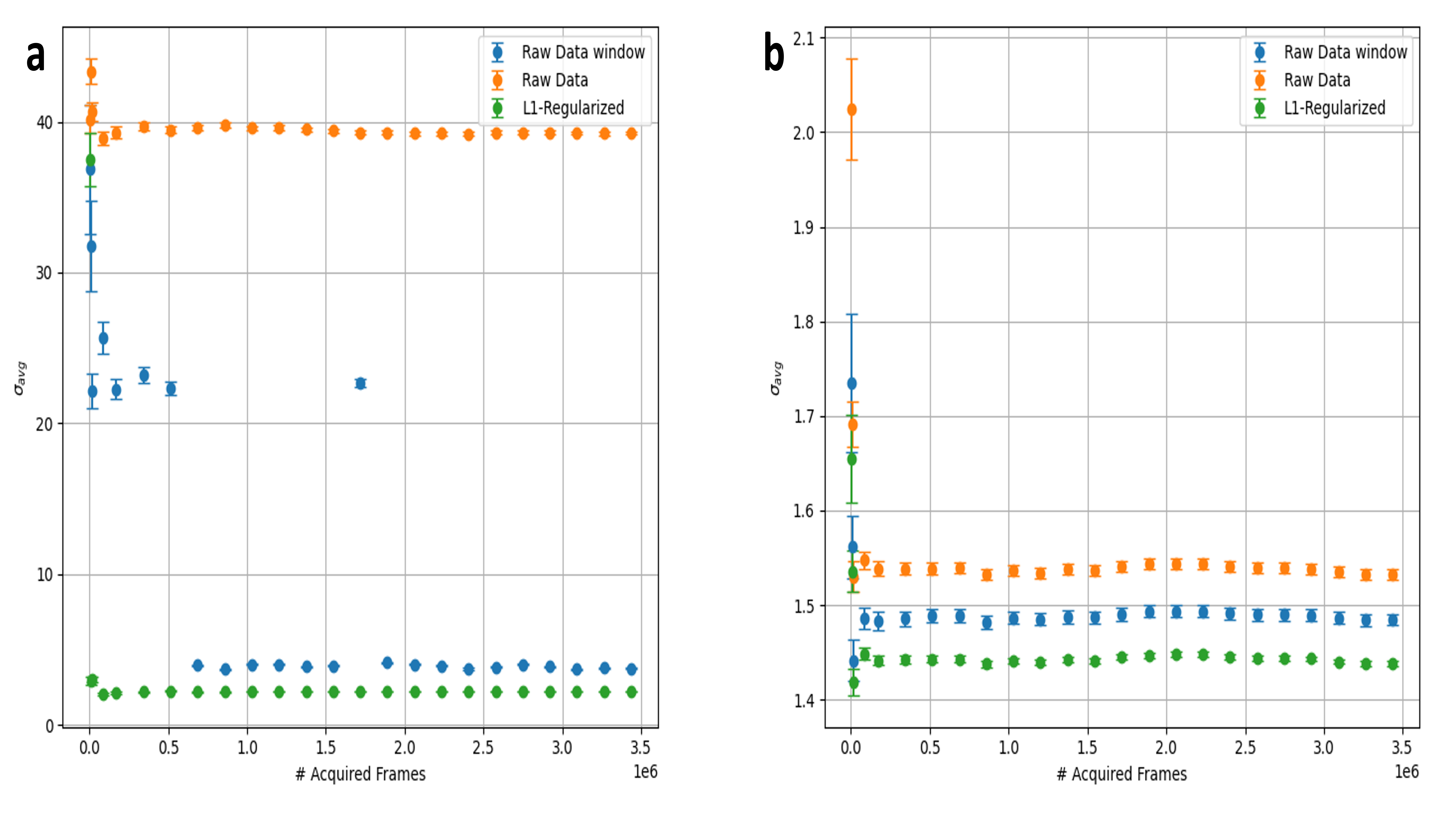}
\caption{Average Gaussian width, $\sigma_{\text{avg}}$, in pixels of the sum-coordinate \textbf{(a)} and minus-coordinate \textbf{(b)} projection of the covariance matrix, as a function of the number of acquired frames.}
\label{fig:sigma_comparison}
\end{figure}

To quantify the spatial correlations, we applied 2D Gaussian fitting to the sum and minus coordinate projections of the covariance matrix distributions. The fitting was performed using the \newline'scipy.optimize.curve\_fit' function in Python, using the model function in Equation~\ref{eq:gaussian}. From the fit parameters, we extracted $\sigma_x$ and $\sigma_y$ (along with their associated uncertainties), and computed the average width $\sigma_{\text{avg}} = (\sigma_x + \sigma_y)/2$. The results are shown in Figures \ref{fig:sigma_comparison}.

\begin{equation}
f(x, y) = A \exp\left[ -\left( \frac{(x - x_0)^2}{2\sigma_x^2} + \frac{(y - y_0)^2}{2\sigma_y^2} \right) \right] + B
\label{eq:gaussian}
\end{equation}

\subsubsection*{Dimension Witness Lower Bound}

We estimated the entanglement dimensionality lower bound using the EPR uncertainty relation:

\begin{equation}
\Delta x \cdot \Delta k = \sigma_x \cdot \sigma_k
\label{eq:epr_product}
\end{equation}

The associated uncertainty is propagated as:

\begin{equation}
\delta(\Delta x \cdot \Delta k) = \sqrt{(\sigma_k \cdot \delta \sigma_x)^2 + (\sigma_x \cdot \delta \sigma_k)^2}
\label{eq:epr_uncertainty}
\end{equation}

\begin{figure}[h]
\centering
\includegraphics[width=0.9\linewidth]{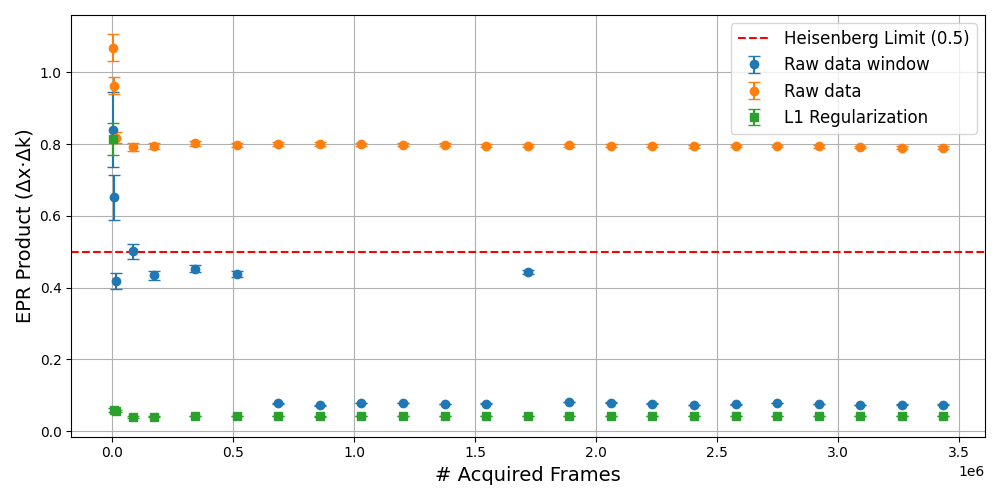}
\caption{EPR as a function of the number of acquired frames.}
\label{fig:setup}
\end{figure}

The dimensionality witness lower bound is then calculated as:

\begin{equation}
d = \frac{1}{e \cdot \sigma_x \cdot \sigma_k}
\label{eq:dim_bound}
\end{equation}

With propagated uncertainty:

\begin{equation}
\delta d = \frac{1}{e} \cdot \sqrt{
\left( \frac{\delta \sigma_x}{\sigma_x^2 \cdot \sigma_k} \right)^2 +
\left( \frac{\delta \sigma_k}{\sigma_x \cdot \sigma_k^2} \right)^2
}
\label{eq:dim_uncertainty}
\end{equation}

\begin{figure}[h]
\centering
\includegraphics[width=0.9\linewidth]{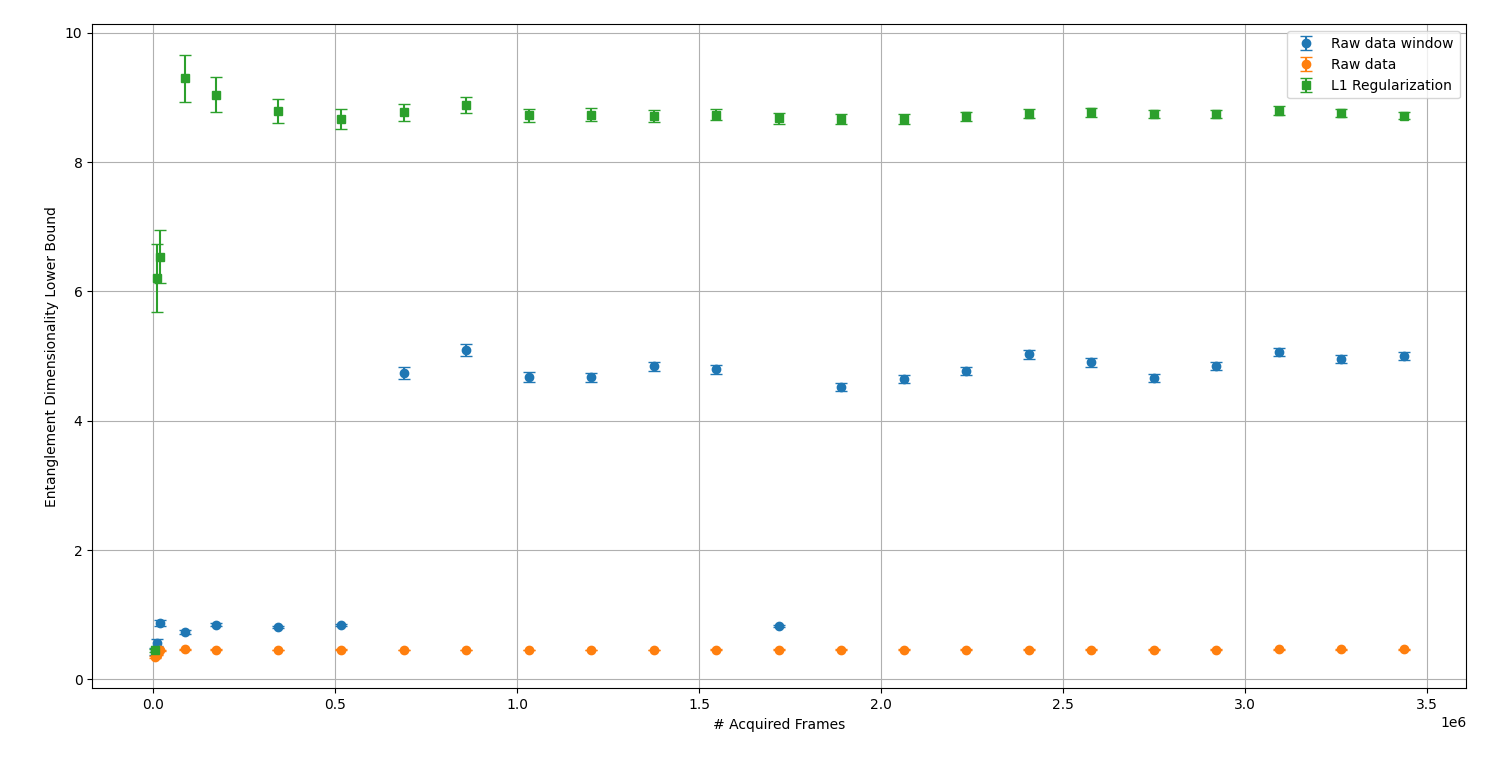}
\caption{Lower bound on entanglement dimensionality as a function of the number of acquired frames.}
\label{fig:setup}
\end{figure}

\subsection*{Hyperparameter Selection}
We empirically tuned the regularization parameter $\lambda$ for the $\ell_1$ optimization to balance signal fidelity and noise suppression. A value of $\lambda = 10^{-4.8}$ with learning rate of $10^{-1.43}$(for EMCCD) yielded optimal SNR without oversmoothing. For SPAD data, no shift was needed.

\printbibliography

\section*{Data Availability}
Data underlying the results presented in this paper are not publicly available at this time but may be obtained from the authors upon reasonable request.

\section*{Code availability}
The codes used to process the data are available from the corresponding authors upon reasonable request.

\section*{Acknowledgment}
The authors thank Raphael Guitter and Baptiste Courme for their guidance and support in learning how to perform the experiments during the stay in Paris, and Bienvenu Ndagano for providing the SPAD camera data used in this study.
This research was supported by the Israel Science Foundation (ISF), Grant No. 3620/24 and the Israel Ministry of Innovation, Science, and Technology, Grant No. 2033419. We acknowledge support from the Russell Berrie Nanotechnology Institute (RBNI) and from the Hellen Diller Quantum Center at the Technion. H.D. acknowledges funding from the ERC Starting
Grant (No. SQIMIC-101039375).

\section*{Funding}
This research was supported by the Israel Science Foundation (ISF), Grant No. 3620/24 and the Israel Ministry of Innovation, Science, and Technology, Grant No. 2033419.

\section*{Author Contributions}
S.L. performed the experiment and analysed the results. S.L., H.D., R.T. and
G.B. conceived and discussed the experiment. G.B. and H.D. provided experimental support with the EMCCD setup. R.T. provided guidance on the algorithm. All authors discussed the results and contributed to writing and revising the manuscript.

\section*{Competing Interests}
The authors declare no competing interests.

\end{document}